# Magnetoelectric energy in electrodynamics: Magnetoelectricity, bi(an)isotropy, and magnetoelectric meta-atoms

E. O. Kamenetskii

Department of Electrical and Computer Engineering,
Ben Gurion University of the Negev, Beer Sheva, Israel

June 19, 2023

**Abstract**

Magnetoelectricity denotes the relationship between electric polarization and magnetization. In materials with an intrinsic magnetoelectric (ME) effect, the energy density comprises the polarization, magnetization and ME energy densities. These three components of energy define local (subwavelength) characteristics of electromagnetic (EM) responses in multiferroic materials. In a subwavelength domain, coupling between the electric and magnetic dipole oscillations forms the ME field structures which are characterized by the violation of both spatial and temporal symmetry. Unlike multiferroics, bi(an)isotropic metamaterials are associated with an EM response characterized only by spatial symmetry breaking. This also applies to chiral materials. Since no "intrinsic magnetoelectricity" is assumed in such structures, any concepts about the stored ME energy are not applicable. This clearly points to the effect of nonlocality. That is why the basic concepts of bi(an)isotropy can only be analyzed by the EM far-field characteristics. In this paper, we argue that in the implementation of local (subwavelength) ME meta-atoms and systems for near-field probing of chirality, the concept on ME energy is crucial. Real ME energy can occur when ME fields in a singular subwavelength domain are characterized by a violation of both the symmetry of time reversal and spatial reflection.

## I. INTRODUCTION

In electrodynamics, we consider electric and magnetic fields that are bundled together. This coupling forms EM fields. Magnetoelectricity is associated with the effect of coupling electric and magnetic dipoles. This effect underlies the properties of ME materials which we can analyze in the dipole approximation. Electrodynamics of ME materials is the extension of the ME effect to the dynamic regime. The electric and magnetic fields bundled together due to coupling between the electric and magnetic dipole oscillations in a subwavelength domain, form ME fields. The energy of resonances in this confined structure is called ME energy. In this paper, we show why the concept of ME energy plays a defining role in electrodynamics of ME media. This concerns electrodynamics of multiferroics, bi(an)isotropic metamaterials, and chiral media. Synthesizing of ME polarizability and ME fields in resonant structures is key aspect for the creation of ME meta-atoms and ME near-field probes.

Magnetoelectricity can be associated with various mechanisms of the relationship between electric polarization and magnetization. The intrinsic ME coupling in multiferroics is due to the electric polarization induced by spatially modulated spin structures. This happens when the ordering of the electron spins breaks the spatial symmetry of the inversion. In multiferroics, the two main symmetries associated with magnetism and ferroelectricity are time reversal symmetry and space reversal symmetries. For EM waves propagating in a medium with intrinsic ME properties, the energy balance equation is relevant if the constitutive parameters satisfy the long-wavelength (quasistatic) limit. That is, for monochromatic waves, these parameters must be physically valid for the wave vector $|\vec{k}| \to 0$.

In the ME electrodynamics, subwavelength fluctuations in the intensity of EM excitation arise only at certain ratios of the amplitudes of the electric and magnetic fields. This relationship defines local dynamics of ME resonances. In a ME subwavelength region where both the electric and magnetic fields exist, energy conversion between average stored energies can occur through a certain circular process of energy exchange. This implies the presence of the subwavelength power-flow circulation, which can be carried out with synchronous rotation in time of the complex-amplitude vectors. In a dynamical regime, the multiferroic sample becomes unstable under EM plane wave radiation. It disintegrates at rotating-field chiral domains with power-flow vortices and axion regions. The field amplitudes are associated with the energy excitations of the vortices.

The effect of energy circulation in the subwavelength ME domain is well modeled based on the analysis of the field structure of quasistatic oscillations in a quasi-2D disk made of a magnetic insulator. ME response of a quasi-2D ferrite disk is characterized by the violation of both parity (*P*) and time-reversal (*T*) symmetry. However, the effect preserved in the combined *PT* symmetry. The modeling of a multiferroic as a composition of rotating-field chiral domains with power-flow vortices can be useful for studies of interaction of ME material with EM radiation. When considering a ferrite resonator with quasistatic oscillations as a model of a building block of the multiferroic structure, we can model such a material composition as an array of quasi-2D ferrite disks.

In a common scenario, bianisotropy is considered as an EM time-even effect, fundamentally related only to the violation of spatial symmetry. Since no "intrinsic bianisotropy" is assumed, any concepts about the stored ME energy are not applicable. In other words, no mechanisms of quasistatic energy of interaction between electric and magnetic dipoles in a subwavelength region can be proposed in this case. This clearly points to the effect of nonlocality: "local coupling" is possible only via EM continuum. For this reason, the basic concepts of bianisotropy are analyzed from the EM far-field characteristics. It is obvious that the use of the near-field technique for studying bianisotropic metamaterials is possible only if the mechanism of local ME energy is proposed. This mechanism presumes the creation of subwavelength resonant structures with coupled quasi-magnetostatic and quasi-electrostatic oscillations. The electric and magnetic fields bundled together in such a resonator form ME fields which are characterized by the eigenstate spectrum of ME energy. In a particular case of a biisotropic medium, the synthesis of ME polarizability and ME fields can be represented as *combined effect* of local responses in Tellegen and chiral media. This is the effect of violation of both spatial inversion symmetry and time reversal symmetry in a subwavelength structure. The physics of such ME resonators underlies the fundamental characteristics of ME meta-atoms for metamaterials and ME near-field probes for testing molecular chirality.

## II. MAGNETOELECTRICITY

### A) Magnetoelectric energy

In the materials where electric polarization and magnetization are treating as the order parameters, the magnetization can be manipulated by electric field, and vice versa. Multiferroics are insulating materials characterized by both magnetic and dielectric orders. In the case of multiferroics, ferroelectric and ferromagnet phenomena are in the same phase and the magnetic properties are coupled to the electric properties through a ME interaction. In these structures, the ME effect is described by an expansion of the free energy. The energy density consists of three parts of the potential energy: the *polarization* energy density, the *magnetization* energy density, and the *magnetoelectric* energy density [1–6].

The extension of the ME effect to the dynamic regime opens a new area in electromagnetism. ME effects in multiferroics can only occur when *both the spatial and the temporal inversion symmetry are broken*. To describe the dynamical processes of the plane wave propagation in such a medium with a linear ME effect, the integral-form constitutive relations (ICR) are used [7]

$$D_i(t,\vec{r}) = (\varepsilon_{ij} \circ E_j) + (\xi_{ij} \circ H_j), \tag{1}$$

$$B_i(t,\vec{r}) = (\zeta_{ij} \circ E_j) + (\mu_{ij} \circ H_j). \tag{2}$$

The integral operators on the right-hand side of these expressions have the form similar to the integral operator:

$$(\varepsilon_{ij} \circ E_j) = \int_{-\infty}^{t} dt' \int d\vec{r}' \varepsilon_{ij}(t,\vec{r},t',\vec{r}') E_j(t',\vec{r}'). \tag{3}$$

The kernels of the operators in the above ICRs are responses of a medium to the $\delta$-function electric and magnetic fields. Convergence of integrals in the ICRs can be proven if one shows a physical mechanism of influence of short-time and short-space *quasistatic* interactions on the medium polarization properties. When a ME medium is time invariant and spatially homogeneous, the ICRs have a temporal and space convolution form. For a temporary dispersive ME medium, we have constitutive relations:

$$\vec{D}(\omega) = \vec{\vec{\varepsilon}}(\omega)\vec{E} + \vec{\vec{\xi}}(\omega)\vec{H}, \qquad \vec{B}(\omega) = \vec{\vec{\zeta}}(\omega)\vec{E} + \vec{\vec{\mu}}(\omega)\vec{H}. \tag{4}$$

In Ref. [8], constitutive relations (4) are used for an analysis of ME resonances in helimagnet. These resonances are expected to be enhanced within materials that support electromagnons – fundamental excitations that exhibit both electric and magnetic dipole moments. Electromagnon excitations in multiferroic materials are skyrmions. The multiferroic nature of the skyrmions is due to spin-induced ferroelectricity [9]. Eqs. (4) are also considered as general-form constitutive relations in axion electrodynamics [10]. In the particular case of scalar constitutive parameters, equations (4) are used to describe the optical activity of natural chiral media in the Condon formalism [11].

In the dipole approximation, for EM waves propagating in a medium with intrinsic ME properties, one can use the energy balance equation:

$$-\nabla \cdot \langle \vec{S} \rangle = \frac{\partial \langle W \rangle}{\partial t} + \langle P \rangle. \tag{5}$$

In this continuity equation for the time-averaged values of periodic functions, $\langle \vec{S} \rangle$ is a Poynting vector, $\langle W \rangle$ is an average stored energy, and $\langle P \rangle$ describes dissipative losses. The energy balance equation is relevant if the constitutive parameters in Eq. (4) satisfy the *long-wavelength (quasistatic) limit*. That is, for monochromatic waves, these parameters must be physically valid for the wave vector $|\vec{k}| \to 0$.

To derive the energy balance equation in temporary dispersive dielectric and magnetic media, one must use the regime of propagation of *quasi-monochromatic* EM waves [4]. For a ME medium, such a quasi-monochromatic behavior was considered in Ref. [12]. The fields are expressed as $\vec{E} = \vec{E}_m(t,\vec{r})\, e^{i(\omega t - \vec{k}\cdot\vec{r})}$ and $\vec{H} = \vec{H}_m(t,\vec{r})\, e^{i(\omega t - \vec{k}\cdot\vec{r})}$, where complex amplitudes $\vec{E}_m(t,\vec{r})$ and $\vec{H}_m(t,\vec{r})$ are time and space smooth-fluctuation functions. It is supposed that in the frequency regions of the transparency of the EM-wave propagation, the concept of an internal-energy density in alternative fields can introduced in the same sense as it is used in *magnetoelectrostatic* structures. For the time-averaged stored energy $\langle W \rangle$ we have [12]:

$$\langle W \rangle = \frac{1}{4}\left\{ \frac{\partial(\omega \varepsilon_{ij}^h)}{\partial \omega} E_i^* E_j + \frac{\partial(\omega \mu_{ij}^h)}{\partial \omega} H_i^* H_j + \frac{\partial\left[\omega(\zeta_{ij}^h + \xi_{ij}^h)\right]}{\partial \omega}\left(H_i^* E_j\right)^h + \frac{\partial\left[\omega(\zeta_{ij}^{ah} - \xi_{ij}^{ah})\right]}{\partial \omega}\left(H_i^* E_j\right)^{ah} \right\}. \qquad (6)$$

For a lossless medium, tensors $\vec{\vec{\varepsilon}}$ and $\vec{\vec{\mu}}$ are Hermitian and $\vec{\vec{\xi}}^h = \vec{\vec{\zeta}}^h$, $\vec{\vec{\xi}}^{ah} = -\vec{\vec{\zeta}}^{ah}$. In a quasistatic limit, $\lambda \to \infty$, the parts of average stored energy in Eq. (6) correspond to potential energies in a multiferroic [1–6]. These are the electric-field energy density $W_E$ and the magnetic-field energy density $W_M$ expressed by the first two terms on the right-hand side of Eq. (6). The ME energy density $W_{ME}$ is expressed by the last two terms on the right-hand side of Eq. (6). In a weakly absorbing medium, the dissipative losses are described by the relation [12]:

$$\langle P \rangle = \frac{1}{2}\omega\left[ \varepsilon_{ij}^{ah} E_i^* E_j + \mu_{ij}^{ah} H_i^* H_j + \left(\zeta_{ij}^h - \xi_{ij}^h\right)\left(H_i^* E_j\right)^{ah} + \left(\zeta_{ij}^{ah} + \xi_{ij}^{ah}\right)\left(H_i^* E_j\right)^h \right]. \qquad (7)$$

Superscripts *h* and *ah* in Eqs. (6), (7) denote, respectively, the Hermitian and anti-Hermitian parts of tensors of the second rank.

It is important to note that the continuity equation (5) is valid only when definite constraints are imposed to slowly time-varying amplitudes of the field components:

$$E_{m_i}^*(t) \frac{\partial H_{m_j}(t)}{\partial t} = H_{m_j}(t) \frac{\partial E_{m_i}^*(t)}{\partial t}. \qquad (8)$$

The physical meaning of these constraints is quite clear. Subwavelength fluctuations in the intensity of EM excitation in a ME medium arise only at certain ratios of the amplitudes of the electric and magnetic fields. The equation (8) can be transformed as follows:

$$\frac{E_{m_i}^*(t)}{H_{m_j}(t)} = \frac{\partial E_{m_i}^*(t)/\partial t}{\partial H_{m_j}(t)/\partial t} = \frac{dE_{m_i}^*(t)}{dH_{m_j}(t)}, \qquad (9)$$

where $dE_{m_i}^*$ and $dH_{m_j}$ are differentials of the corresponding fields.

The constraints (9) can be considered as certain symmetry conditions for the field structure. The derivation of the continuity equation (5) from symmetry principles of Eq. (9) is related to Noether's theorem. By virtue of Noether's theorem, the invariance of the action under translation in time, translation in space, and rotation, implies the existence of the conservation of energy, linear momentum, and angular momentum, respectively. Noether's theorem is used to investigate symmetries and related conserved quantities in Maxwell's equations [13]. With Noether's theorem, we demonstrate electric–magnetic symmetry in ME electromagnetism. It becomes obvious that *without constraints (9) it is impossible to use the concept of ME energy*. Eq. (9) implies that there exists a linear time-relation coupling between complex amplitudes of the fields. In a general form, we can write [14]

$$E_{m_i}(t) = T_{ij} H^*_{m_j}(t). \qquad (10)$$

The matrix $[T]$ is a field-polarization matrix. This is an invariant defined for a specific type of ME medium. Components of matrix $[T]$ are complex quantities. To find parameters of the polarization matrix $[T]$, one should solve an electromagnetic boundary problem with the known medium constitutive parameters. When the parameters of matrix $[T]$ are found, an average energy density in a medium can be determined. In general, we have an integro-differential problem.

Being not electromagnetic in nature, the energy density $W_{ME}$ plays a special role in forming a topological structure of the fields. We know that for any EM process in a lossless non-ME medium $|\langle W_E \rangle| = |\langle W_M \rangle|$. For a propagating monochromatic plane wave, the electric energy density and magnetic energy density are equal to each other at every instant of time. For an EM resonator (such as, for example, a closed metal-wall cavity or a dielectric resonator), we have a quasistatic process when electrical energy is converted into magnetic energy and vice versa over a time period of the EM radiation (Fig. 1). "Mediators" of these transformations are electric conduction currents and/or electric currents of polarization (displacement currents). Now, in electrodynamical ME effects in multiferroics, we have the three parts of the average stored energy, $\langle W_E \rangle$, $\langle W_M \rangle$, and $\langle W_{ME} \rangle$. In this case, it should be assumed that there exists also a specific "mediator" associated with the ME effect, which mutually converts the electric and magnetic energies. In a subwavelength region where both the electric and magnetic fields exist, energy conversion between average stored energies $\langle W_E \rangle$, $\langle W_M \rangle$, and $\langle W_{ME} \rangle$ can occur through a certain *circular process of energy exchange*. This implies the presence of a *power-flow circulation*, which can be carried out with *synchronous rotation* in time of both complex-amplitude vectors $E_{m_i}$ and $H_{m_j}$ [12, 14]. So, the process is accompanied by *power-flow vortices in subwavelength domains*. The power flows of the circulating adiabatic processes of the energy interchange are not exclusively the EM power flows. Since in multiferroics the temporal inversion symmetry is broken, we can observe the right- and left-hand vortices (Fig. 2).

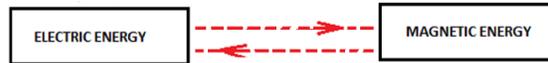

**Fig. 1.** EM resonator. In a lossless resonator $W = W_{el} + W_{magn}$. On the scale of the EM wavelength, a transformation of energy takes place: $W_{el} \leftrightarrow W_{magn}$.

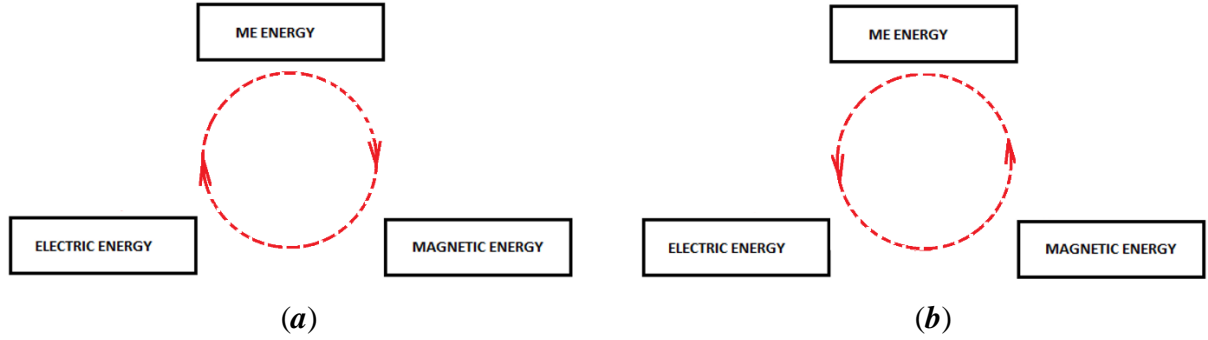

**Fig. 2.** ME resonator. Right-handed (*a*) and left-handed (*b*) power-flow vortices in subwavelength domains. In a lossless ME resonator $W = W_{el} + W_{magn} + W_{me}$.

We have a ME "trion"– a localized (subwavelength) resonant excitation with energies of three subsystems (electric, magnetic, and magnetoelectric). The energy states of ME "trions" can be split in an applied magnetic field. For two opposite directions of a normal bias field, there are two types of ME "trions" with different chirality (Fig. 3).

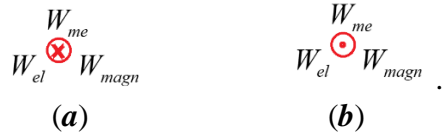

**Fig. 3.** (*a*), (*b*) Two types of ME "trions" with different chirality for two opposite directions of a normal bias field.

**B) Subwavelength magnetoelectric domains**

In multiferroics, the ME coupling is manifested by the formation and interactions of domains and domain walls. It such domain structures, topological vortices can occur [15]. With that, the building blocks for multiferroicity – the ME monopole, toroidal moment, and quadrupole – are related respectively to the isotropic diagonal, antisymmetric off-diagonal and traceless components of the ME tensor. [16]. Our analysis of ME domains is related to ME electrodynamics. In quasistatic processes of energy circulation in a subwavelength-domain region, the electric and magnetic fields are, in general, neither mutually perpendicular nor mutually parallel. The fields are mutually parallel in the so-called axion area [17]. In the area where the electric and magnetic fields are mutually perpendicular, we have a maximum of the power flow. It was shown [18] that the macroscopic order effects in multiferroics enable the coupling between the axion and matter fields. In this case, the electric and magnetic dipoles can be induced without using any external magnetic or electric fields due to spontaneously broken symmetries. The field structures of subwavelength domains are associated, respectively, with the regions of antisymmetric (toroidal) and monopolar-contribution ME responses observed in multiferroics [19 – 22]. The presence of such toroidal domains in a static limit is correlated with spin orderings with persistent orbital magnetic currents.

In a dynamical regime, the multiferroic sample becomes unstable under EM plane wave radiation. It disintegrates at rotating-field chiral domains with power-flow vortices and axion regions. The field amplitudes are associated with the energy excitations of the vortices. Symmetries are broken by the vortex solutions. The vortex solutions minimize free energy. The space period of domain textures is incommensurate with the space period of crystal structures of a multiferroic material. This justifies the use of a continuum approximation. Another question is about the time period of rotation of complex-amplitude vectors. The direction of an orbital angular momentum of the fields in domains does not relate to the direction of polarization of the incident EM field. It means that when the phase of the incident EM field is shifted on $\pi$, orbital rotation of complex-amplitude vectors is at $2\pi$. So, for the frequency $\omega$ of the incident EM field, we have orbital rotation of complex-amplitude vectors with the frequency $\Omega = 2\omega$. Orbital angular momentum of domains is half-integer. This can be possible due to topological effects in subwavelength domains. For EM radiation, these domains are topological polarization singularities.

When propagation vector of EM radiation is along $z$ axis, the transversal-field structure with complex-amplitude vectors $E_{m_\perp}$ and $H_{m_\perp}$ rotate synchronically on a $xy$ plane of a subwavelength domain. In this case, Eq. (9) is fulfilled when

$$\frac{dE_{m_\perp}}{dH_{m_\perp}} = \frac{d\varphi E_{m_\perp}}{d\varphi H_{m_\perp}} = \frac{E_{m_\perp}}{H_{m_\perp}}, \quad \Omega = \frac{d\varphi}{dt}. \tag{11}$$

This situation is illustrated in Fig. 4. There can be clockwise and counterclockwise rotations of the fields. These rotations can be observed separately when a bias magnetic field is applied.

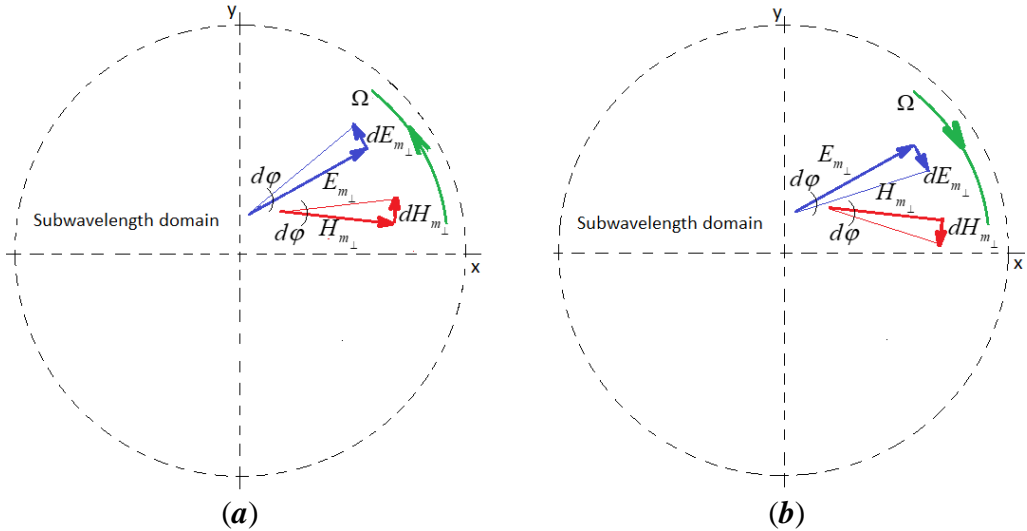

(*a*)          (*b*)

**Fig. 4.** Rotating fields in a chiral domain. Transversal-field structure with complex-amplitude vectors $E_{m_\perp}$ and $H_{m_\perp}$ in a given point of a subwavelength ME domain. At the frequency $\omega$ of the incident EM field, the fields in the domain rotate *synchronically* on the *xy* plane with frequency $\Omega = 2\omega$. Orbital angular momentum is half-integer.

It is worth noting the following. When observing rotating-field chiral domains in multiferroics, it is necessary to distinguish between the concepts of angular momentum and pseudoangular momentum. The conservation of angular momentum and the conservation of pseudoangular momentum are two different types of conservation laws. While the conservation of angular momentum follows from invariance under rotations of all the constituents of a solid, the conservation of pseudoangular momentum follows from invariance under rotations of fields while keeping the solid fixed [23]. In our case, we are talking about pseudoangular momentum.

In multiferroic media, the constitutive parameters are tensors. A general analysis of the energy quadratic relations and the circulation process of the energy in a ME domain is quite complicated. We, however, have the opportunity to observe angular momenta of the fields in ME chiral domains by modeling the circulation process of the energy in a small (subwavelength) ferrite resonator with quasistatic oscillations. Insulating ferrites such as yttrium iron garnet (YIG) play an important role in modern spintronic and magnonic devices. In these isotropic magnetic dielectrics, nontrivial magnetic textures in samples of special geometry can cause the dynamical effect characterized by violation of both spatial inversion and time reversal symmetries. The ME effect observed in a normally magnetized quasi-2D ferrite disk can be described by *scalar* cross-polarization parameters. An analysis of a mesoscopic effect of dynamical magnetoelectricity in such a *subwavelength* sample may be important for further modeling the properties of ME domains in multiferroics.

Suppose that in a lossless structure, the cross-polarization parameters are *scalars* represented as $\xi = \eta - i\gamma$ and $\zeta = \eta + i\gamma$. In this case, for the density of average ME energy we have from Eq. (6):

$$\langle W_{ME} \rangle = \frac{1}{2} \frac{d(\omega\eta)}{d\omega} \text{Re}\left(\vec{E} \cdot \vec{H}^*\right) + \frac{1}{2} \frac{d(\omega\gamma)}{d\omega} \text{Im}\left(\vec{E} \cdot \vec{H}^*\right). \tag{12}$$

Based on this equation, we will analyze the ME properties in a normally magnetized quasi-2D ferrite disk with *quasistatic oscillations*. The structure of the electric and magnetic fields in such a sample was studied numerically and analytically in Ref. [24, 25]. Schematically, it can be modeled as it is shown in Fig. 5. In a central region of the disk, we have only the in-plane field components of a magnetic dipole and an electric quardupole. Assuming the counterclockwise rotation of the fields in this region, we have for transversal electric and magnetic fields lying in the *xy* plane:

$$\vec{E} = \vec{E}_\perp = a_\perp(\hat{\vec{x}} + i\hat{\vec{y}})$$
$$\vec{H} = \vec{H}_\perp = b_\perp(\hat{\vec{x}} + i\hat{\vec{y}}) \tag{13}$$

where $a_\perp, b_\perp$ are complex amplitudes and $\hat{\vec{x}}, \hat{\vec{y}}$ are unit vectors. Suppose that $a_\perp = |a_\perp|$ and $b_\perp = |b_\perp|e^{i\vartheta}$, where $\vartheta$ is an arbitrary angle within $0 \leq \vartheta \leq 90°$. It is a time shift between complex amplitudes of the electric and magnetic fields. A scalar product of the electric-field vector and the complex conjugate magnetic-field vector gives:

$$\left(\vec{E}_\perp \cdot \vec{H}_\perp^*\right)_{xy} = |a_\perp||b_\perp|e^{-i\vartheta}(\hat{\vec{x}} + i\hat{\vec{y}}) \cdot (\hat{\vec{x}} - i\hat{\vec{y}}) = 2|a_\perp||b_\perp|[\cos\vartheta - i\sin\vartheta]. \tag{14}$$

At the same time, for the vector product we have:

$$\left(\vec{E}_\perp \times \vec{H}_\perp^*\right)_{xy} = |a_\perp||b_\perp|e^{-i\vartheta}(\hat{\vec{x}} + i\hat{\vec{y}}) \times (\hat{\vec{x}} - i\hat{\vec{y}}) = -2|a_\perp||b_\perp|[\sin\vartheta + i\cos\vartheta]\hat{\vec{z}}, \tag{15}$$

where $\hat{\vec{z}}$ is a unit vector along *z* axis. The following relation is obvious for any angle $\vartheta$:

$$\text{Im}\left|\vec{E}_\perp \cdot \vec{H}_\perp^*\right|_{xy} = \text{Re}\left|\left[\vec{E}_\perp \times \vec{H}_\perp^*\right]_z\right|, \tag{16}$$

where the term $\text{Re}\left|\left[\vec{E}_\perp \times \vec{H}_\perp^*\right]_z\right|$ is considered as a power flow density along $z$ axis. Since at the center of the disk, we have the structure of a magnetic dipole and an electric quadrupole, the power flux-densities along the z-axis in the upper and lower halves of the disk cancel each other out. It means that in the expression for the average stored energy (12), the term $\text{Im}\left|\vec{E}_\perp \cdot \vec{H}_\perp^*\right|$ must be omitted. The spin-angular-momentum rotation of the fields is characterized by *real ME energy* $\text{Re}\,\vec{E}_\perp \cdot \vec{H}_\perp^* = 2|a_\perp||b_\perp|$ in the disk center. It is important to note that this is a *pseudoscalar*: it changes sign on mirror reflection over $z$ axis. Since this quantity also changes sign when the direction of the bias magnetic field changes, we have *PT symmetry of the ME energy*.

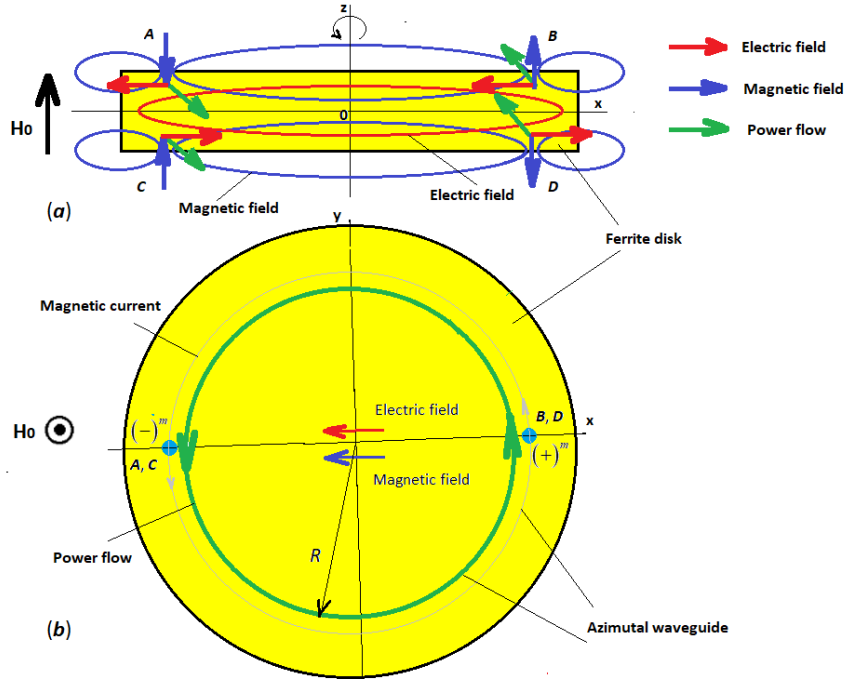

**Fig. 5.** Modeling of energy circulation in the subwavelength ME domain based on an analysis of the field structure in a normally magnetized quasi-2D ferrite disk with quasistatic oscillations. (*a*) Cross-section side view of the field structure in a quasi-2D ferrite disk. (*b*) Top view of the upper plane of the disk. The spin-angular-momentum rotation of parallel electric and magnetic fields in a central region. In a peripheric region, one observes an orbital rotation of "magnetic charges" and a power flow vortex.

On the periphery of a quasi-2D ferrite disk, certain regions are visible. In Fig. 5 these regions are labeled *A*, *B*, *C*, and *D*. Seen from above and below in a vacuum, the regions look like magnetic monopoles. We observe a Gilbert magnetic dipole composed of a magnetic monopole-antimonopole pair. Magnetic monopoles with "magnetic charges" $(+)^m$ and $(-)^m$ rotate at frequency $\Omega = 2\omega$ clockwise or counterclockwise depending on the direction of the bias magnetic field, where $\omega$ is the frequency of EM radiation.

Energy circulation in the subwavelength ME region is the rotational superradiance effect [26]. Generation of waves by a rotating body is a well-known phenomenon [27]. If the obstacle is rotating, waves can be amplified in the process, extracting energy from the scatterer. In our case, incident EM waves are amplified by extracting energy – ME energy – from the collective motion of orbitally rotating magnetic dipoles. At the magnetic-dipolar-mode (MDM) resonances, transfer between angular momenta of a magnetic insulator and an external microwave structure demonstrates generation of vortex flows with a fixed handedness. The field structure of emitted photon has a unique topology. Experimentally, the rotational superradiance in a quasi-2D ferrite disk was analyzed due to the friction effect at the with MDM resonances [28, 29].

The shown in Fig. 5(*a*) transversal electric and magnetic fields lying in the *xz* plane rotate synchronically in regions *A*, *B*, *C*, and *D*. Each of these regions constitutes an azimuthal chiral waveguide. The integral structure indicates the presence of power flow vortices. Assuming that between complex amplitudes of the transversal electric field $\vec{E}_\perp$ and magnetic field $\vec{H}_\perp$ in the *xz* plane there is an arbitrary angle within $0 \leq \theta \leq 90°$, we have an equation like to Eq. (16):

$$\mathrm{Im}\left|\vec{E}_\perp \cdot \vec{H}_\perp^*\right|_{xz} = \mathrm{Re}\left|\left[\vec{E}_\perp \times \vec{H}_\perp^*\right]_y\right|. \tag{17}$$

This relation is relevant for any angle $\theta$. If we assume that at points *A*, *B*, *C*, and *D* the azimuthal coordinate is directed along or against the *y*-axis, then we have azimuthal energy circulation. The real power flows along the azimuthal coordinate are accompanied by the imaginary parts of the ME energy. For waves propagated in azimuthal waveguides, phase velocity $v_{phase} = \Omega R > v_{ferrite}$, where $v_{ferrite}$ is the velocity of the EM wave propagation in a ferrite material. A simple calculation shows that in the YIG disk used in the microwave experiments [28, 29], $v_{phase}$ exceeds $v_{ferrite}$ by about two orders of magnitude. An inequality $v_{phase} > v_{ferrite}$ is the condition of rotational supperradiance. At the same time, the azimuthal wave propagation is with the group velocity $v_{group} \ll v_{ferrite}$.

For our analysis of the power-flow circulation in a subwavelength ME domain, a certain analogy with the EM wave propagation in a metallic-wall lossless waveguide can be useful. In a dispersion characteristic of this waveguide, shown in Fig. 6 (*a*), we see that at the cut-off frequency, an angle between the transversal components of the electric and magnetic fields is $\theta = 90°$. In this point, we have $v_{phase} = \infty$ and $v_{group} = 0$. The angle $\theta = 0°$ corresponds the asymptotic when $v_{phase} = v_{group} = c$. The red spot on the dispersion curve indicates the situation when $0 < \theta < 90°$. In Fig. 6 (*a*), we can see also that in a case of our azimuthal chiral waveguide, we have $\mathrm{Re}\left|\left[\vec{E}_\perp \times \vec{H}_\perp^*\right]_y\right| = 0$ for $\theta = 0°$. There $v_{phase} = \infty$ and $v_{group} = 0$ for this angle. We have $v_{phase} = v_{group} = v_{ferrite}$ for $\theta = 90°$. The red spot on the dispersion curve of the azimuthal chiral waveguide also indicates the situation when $0 < \theta < 90°$. For better understanding this analogy, the *k*-space picture of the macroscopic, mesoscopic, and microscopic scales of oscillations for a certain frequency region is shown in Fig. 6 (*b*). In this figure, EM waves with $\lambda_{em}$ are the waves, which are described by macroscopic Maxwell equations. Quasistatic waves with $\lambda_{qs}$ mean mesoscopic quasistatic oscillations in a subwavelength sample. Atomic-scale waves with $\lambda_{at}$ mean microscopic electron-wave oscillations. We have $\lambda_{em} \gg \lambda_{qs} \gg \lambda_{at}$. Quasistatic oscillations in a 3D-confined subwavelength sample are viewed as processes occurring in the quasistatic limit $\lambda_{em} \to \infty$.

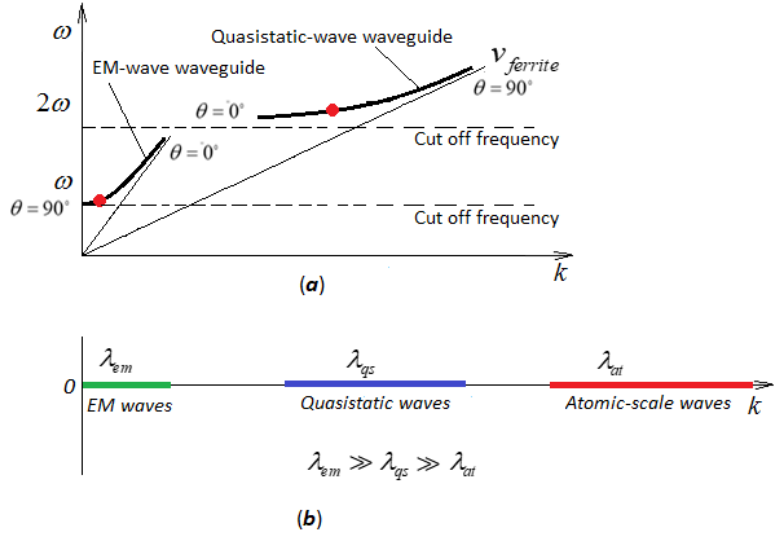

**Fig. 6.** Illustration of internal wave processes in a subwavelength ME domain. (*a*) Dispersion characteristics of a metallic-wall EM-wave waveguide and a quasistatic-wave azimuthal waveguide. (*b*) The *k*-space picture of the macroscopic, mesoscopic, and microscopic scales of oscillations.

When the ME domain interacts with an external EM radiation, a singularity is observed in the center of a vortex configuration. The gyrotropic motion leads to the twisting of the core after it has been excited. In the modeling of a ME domain by a subwavelength ferrite resonator, this effect can be viewed as shown in Fig. 7.

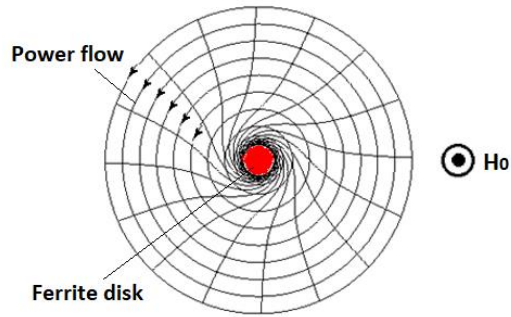

**Fig. 7.** For external EM radiation, subwavelength ME domains are topological polarization singularities. For a given direction of a bias magnetic field, the outflow of EM radiation occurs with the same rotation of the power flow as the incoming radiation. EM radiation can escape the subwavelength domain because it travels through a ferrite disk at a speed less than the speed of light in free space.

Ferroelectricity and magnetism are strongly coupled. Magnetic order causes ferroelectricity and therefore giant ME effect. Our modeling of a multiferroic as a composition of rotating-field chiral domains with power-flow vortices and axion regions can be useful for studies of interaction this material with EM radiation. When considering a ferrite resonator with quasistatic oscillations as a model of a building block of the multiferroic structure, we do not assume here that dynamical magnetoelectricity in this material is due to magnetoelastic and piezoelectric effects. We can model such a composition as an array of quasi-2D ferrite disks. This structure of nine disks placed in a microwave waveguide is shown in Fig. 8. This is a result of our numerical studies in Ref. [30].

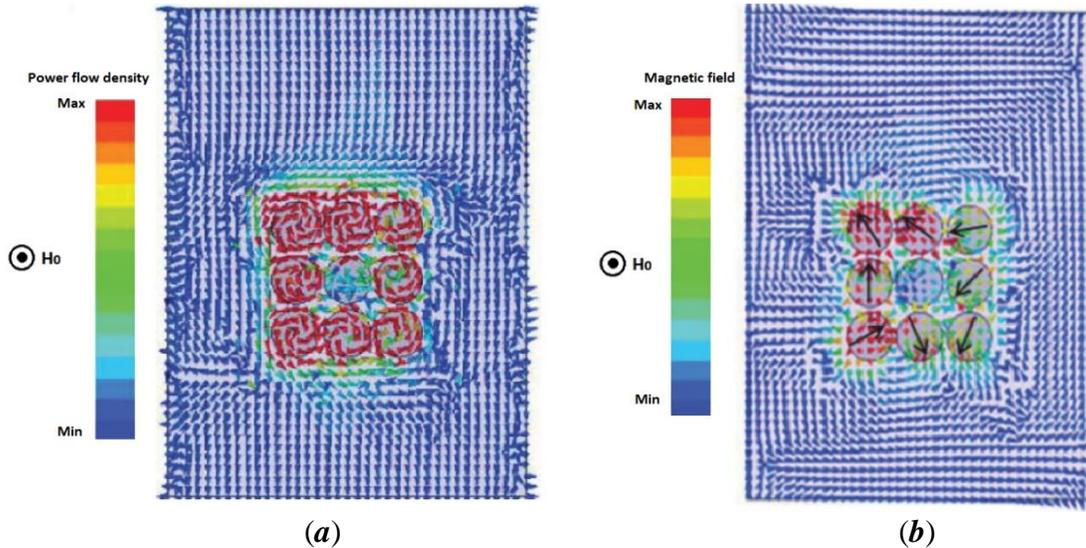

(*a*)          (*b*)

**Fig. 8.** Modeling of a multiferroic structure in the form of an array of subwavelength quasi-2D ferrite disks. The power-flow distribution (*a*) and the magnetic-field distribution (*b*) for a nine-particle array with a center of symmetry.

The ME effect observed in a normally magnetized quasi-2D ferrite disk is *PT* symmetric. It is, however, a *local PT* symmetry of an isolated ME building block. When this sample is placed inside a waveguide or cavity, the *PT* symmetry is broken. For a multiferroic structure presented as a composition of subwavelength ME domains, a *global PT* symmetry violation occurs. This is a ME polaritonic structure. For a given direction of a bias magnetic field, we observe a counterclockwise twisted photons in vacuum above and below the disks. This is an *orbital-angular-momentum* (OAM) field pattern. Fig. 9 shows the OAM array generation in the vacuum region of a microwave waveguide.

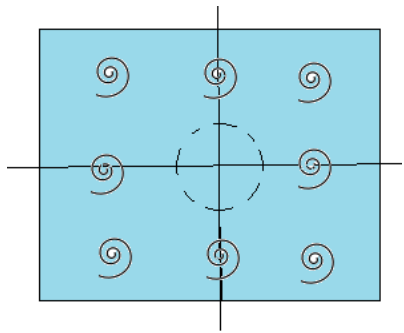

**Fig. 9.** OAM field pattern generated in the vacuum region of a microwave waveguide by nine quasi-2D ferrite disks. At the center of this pattern the field intensity is zero.

Summarizing the basic concepts analyzed in this section, we should note the following. The concept of ME energy is of decisive importance in the analysis of the interaction of multiferroics with EM radiation. This notion can be introduced only if certain constraints are imposed on the field structure. From these constraints, it follows that in a dynamical regime, the multiferroic sample disintegrates at rotating-field chiral domains. The rotations of fields, while keeping the solid fixed, result in the appearance of clockwise and counterclockwise twisted photons outside the sample. These special-form twisted photons, called ME photons, are characterized by the electric-field energy, magnetic-field energy, and ME-field energy.

### III. BI(AN)ISOTROPY

## A) Bianisotropy

In multiferroics, the two basic symmetries associated with magnetism and ferroelectricity are time-reversal symmetry and space-inversion symmetry. In quadratic relations considered above for multiferroics, we have terms of the type $v_{ij}E_iH_j$. In these terms, the tensor components $v_{ij}$ must change a sign upon $\vec{r} \to -\vec{r}$ and $t \to -t$ [4, 19 – 22]. In a medium where spatial inversion and time reversal symmetries are both broken, magnetic and electric fields couple in a way that is fundamentally different from the EM process described by Maxwell's equations.

In a common scenario, bianisotropy is fundamentally related to a response characterized by *only the violation of spatial symmetry*. In Maxwell's equations at frequency $\omega$ written as

$$i\begin{pmatrix} 0 & \vec{\nabla}\times \\ -\vec{\nabla}\times & 0 \end{pmatrix}\begin{pmatrix} \vec{E} \\ \vec{H} \end{pmatrix} = \omega\begin{pmatrix} \vec{\vec{\varepsilon}} & \vec{\vec{\xi}} \\ \vec{\vec{\zeta}} & \vec{\vec{\mu}} \end{pmatrix}\begin{pmatrix} \vec{E} \\ \vec{H} \end{pmatrix}, \tag{18}$$

bianisotropic response is considered as an EM *time-even* effect. Despite the fact that in this case constitutive relations may have the form of Eq. (4), they are not considered as obtained based on ICR (1), (2) with the quasistatic interactions on the medium polarization properties. Since it is assumed that the ME susceptibility components change sign only upon spatial inversion, no idea about the *ME energy* in bianisotropic response can be accepted.

Generally, in the dipole approximation, bianisotropic meta-atoms are considered as a structure of electric-dipole and a magnetic-dipole moments, which are supposedly locally coupled in a 3D-confined subwavelength region. Oscillating electric dipoles are building blocks for electric source. A small current loop can be viewed as a magnetic dipole as long as one is kept away from penetrating the loop. This model essentially consists of neglecting the spatial variation of the electromagnetic field over the meta-atom – instead, one uses the field at the location of the element [31, 32]. The main idea of bianisotropy was to generalize constitutive relations in macroscopic electrodynamics. Bianisotropy is fully connected to the geometrical structure of meta-atoms and caused exclusively by the absence of their inversion symmetry. Since no "*intrinsic bianisotropy*" is assumed, no mechanisms of quasistatic energy of interaction between small electric and magnetic dipoles can be proposed in this case. This clearly points to the effect of nonlocality in such an EM structure: "local coupling" is possible only via EM continuum. Electromagnetically, the local (without the retardation effects) cross-coupling between the electric and magnetic polarizations in a subwavelength 3D region is impossible. Maxwell equations do not describe coupling between the quasimagnetostatic (in neglect of an electric displacement current) and quasielectrostatic (in neglect of a magnetic displacement current) solutions [4, 33]. For this reason, the approach based on scattered *far-field probing* is mainly utilized [34 – 38]. In the dipole approximation, the extinction power, which shows how meta-atom modifies and suppresses the incident wave, calculated as [39, 40]

$$P = -\frac{\omega}{2} \text{Im}\left(\vec{p} \cdot \vec{E}^* + \vec{m} \cdot \vec{H}^*\right), \tag{19}$$

is not a near-field parameter. It turns out to be erroneous to use a circuit-model analysis of the "potential ME energy" based on effective constitutive parameters of bianisotropic meta-atoms detected by the *propagation-wave* methods [41, 42]. In condensed matter physics of dynamical processes, potential energy is found as an average stored energy in a *quasistatic limit* $\lambda \to \infty$ [4]. When the basic concepts of bianisotropy are analyzed from the EM far-field characteristics (without noting about any mechanism of the energy of local coupling between the electric and magnetic dipoles), we just have only "far-field illusion of the *intrinsic magnetoelectricity*" [43]. It becomes obvious that when studying ME energy, it is necessary to use the near-field technique which is based on the effects of violation of both spatial inversion symmetry and time reversal symmetry.

There is a common consensus [44 – 46] that, in the dipole approximation, *local ME characteristics* of bianisotropic meta-atoms can be retrieval from parameters of a small dipole antenna illuminated by a time-harmonic plane wave. However, when analyzing a free-space wave impedance vs. a distance from a source, we have only a wave impedance in the near-field of E-field antenna or a wave impedance in the near-field of H-field antenna [47]. For a small antenna ($\frac{2\pi}{\lambda} a \ll 1$, where $a$ is the characteristic size of the antenna), there is no wave impedance in the near-field of the "ME-field antenna". In a near-field region, one can observe only the densities of the quasistatic electric-energy $W_E$ (for the E-field antenna) or quasistatic magnetic-energy $W_M$ (for the H-field antenna), but not the density of the quasistatic ME energy, $W_{ME}$. No mechanism of quasistatic energy of interaction between small electric and magnetic dipoles is suggested in this model. The detected "ME characteristics" depend only on the geometry of the bianisotropic meta-atoms. This effect is illustrated in Fig. 10. The "magnetoelectricity" of bianisotropic material is due to an extreme capacitive coupling and extreme inductive coupling between building blocks. These are not the near-field effects of the intrinsic ME interactions between separate bianisotropic meta-atoms. Since a bianisotropic meta-atom is an efficient subwavelength resonator, the "ME modifications" of these sources can be observed only in a far-field region.

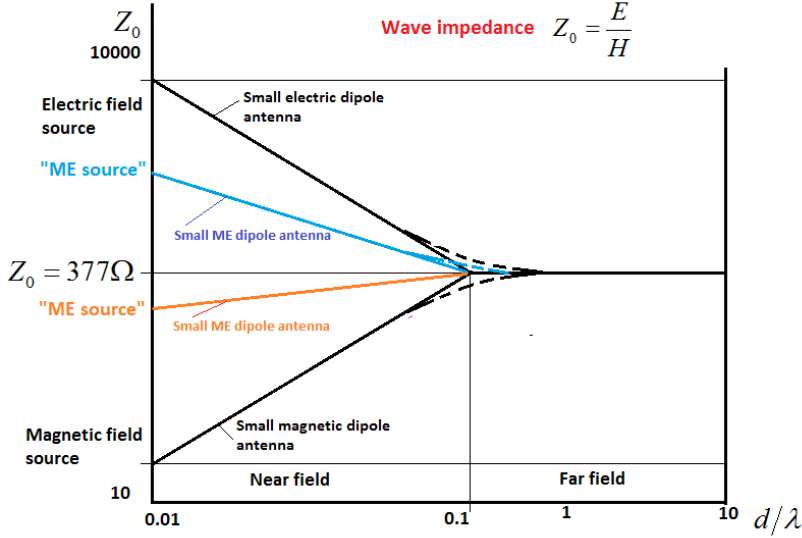

**Fig. 10.** Wave impedance vs. a distance from a source. Meta-atoms as small ME dipole antennas ($ka \ll 1$) are viewed as a local electric or magnetic source of EM radiation. The "ME characteristics" are fully connected to the geometrical structure of meta-atoms. Since a bianisotropic meta-atom is an efficient subwavelength resonator, the "ME modifications" of the electric- on magnetic-field near-field sources can be observed in a far-field region. In a near-field region, one can observe only the quasistatic electric-energy (for the E-field antenna) or quasistatic magnetic-energy (for the H-field antenna), but not the density of the quasistatic ME energy.

## B) Biisotropy

Within the framework of the dipole approximation, biisotropy can be formally considered as a particular case of bianisotropy which is described by scalar constitutive relations:

$$\vec{D}(\omega) = \varepsilon(\omega)\vec{E} + \xi(\omega)\vec{H}, \quad \vec{B}(\omega) = \zeta(\omega)\vec{E} + \mu(\omega)\vec{H}. \tag{20}$$

Following commonly used designation [32, 48], we introduce the constitutive parameters as

$$\varepsilon = \varepsilon_0 \varepsilon_r, \quad \mu = \mu_0 \mu_r, \quad \xi = \sqrt{\varepsilon_0 \mu_0}\,(\chi - i\kappa), \quad \zeta = \sqrt{\varepsilon_0 \mu_0}\,(\chi + i\kappa). \tag{21}$$

We also assume that, generally,

$$\varepsilon_r = \varepsilon_r' + i\varepsilon_r'', \quad \mu_r = \mu_r' + i\mu_r'', \quad \chi = \chi' + i\chi'', \quad \kappa = \kappa' + i\kappa''. \tag{22}$$

In classical electrodynamics of biisotropic media, $\chi$ is called the Tellegen parameter and $\kappa$ is called the parameter of chirality. With the Lorentz reciprocity, the Tellegen medium is considered as a nonreciprocal medium, while the chiral medium as reciprocal. It is assumed that ME responses in axion electrodynamics can be analyzed based on the Tellegen parameters [10, 49 – 51]. In the case of materials with bio-molecular chirality, ME responses are characterized by chiral constitutive relations [11, 52, 53].

For a lossless Tellegen media, we have from Eq. (6):

$$\begin{aligned}\langle W\rangle_{Tellegen} &= \langle W_E\rangle + \langle W_M\rangle + \langle W_{ME}\rangle = \\ &\frac{1}{4}\varepsilon_0 \frac{d(\omega\varepsilon'_r)}{d\omega}\vec{E}\cdot\vec{E}^* + \frac{1}{4}\mu_0 \frac{d(\omega\mu'_r)}{d\omega}\vec{H}\cdot\vec{H}^* + \frac{1}{2}\sqrt{\varepsilon_0\mu_0}\frac{d(\omega\chi')}{d\omega}\operatorname{Re}\left(\vec{H}^*\cdot\vec{E}\right)\end{aligned} \quad (23)$$

For a lossless chiral medium:

$$\begin{aligned}\langle W\rangle_{chiral} &= \langle W_E\rangle + \langle W_M\rangle + \langle W_{ME}\rangle = \\ &\frac{1}{4}\varepsilon_0 \frac{d(\omega\varepsilon'_r)}{d\omega}\vec{E}\cdot\vec{E}^* + \frac{1}{4}\mu_0 \frac{d(\omega\mu'_r)}{d\omega}\vec{H}\cdot\vec{H}^* + \frac{1}{2}\sqrt{\varepsilon_0\mu_0}\frac{d(\omega\kappa')}{d\omega}\operatorname{Im}\left(\vec{H}^*\cdot\vec{E}\right)\end{aligned} \quad (24)$$

We can see that $\langle W_{ME}\rangle \neq 0$ in the case of the Tellegen medium, for a linearly polarized EM wave, and in the case of the chiral medium for a circularly polarized wave. For dissipation losses in Tellegen and chiral media, we have from Eq (7):

$$\langle P\rangle_{Tellegen} = \omega\left[\frac{1}{2}\varepsilon_0\varepsilon''_r\vec{E}\cdot\vec{E}^* + \frac{1}{2}\mu_0\mu''_r\vec{H}\cdot\vec{H}^* + \sqrt{\varepsilon_0\mu_0}\chi''\operatorname{Re}\left(\vec{H}^*\cdot\vec{E}\right)\right], \quad (25)$$

$$\langle P\rangle_{chiral} = \omega\left[\frac{1}{2}\varepsilon_0\varepsilon''_r\vec{E}\cdot\vec{E}^* + \frac{1}{2}\mu_0\mu''_r\vec{H}\cdot\vec{H}^* - \sqrt{\varepsilon_0\mu_0}\kappa''\operatorname{Im}\left(\vec{H}^*\cdot\vec{E}\right)\right]. \quad (26)$$

In general, it can be stated that for a biisotropic medium and an *elliptically polarized* plane wave, there are *joint chiral and Tellegen responses* with violation of *both spatial and temporal inversion symmetries*. The ME energy and the ME dissipation losses are:

$$\langle W_{ME}^{biisotr}\rangle = \langle W_{ME}^{Tell}\rangle + \langle W_{ME}^{chir}\rangle \quad (27)$$

and

$$\langle P_{ME}^{biisotr}\rangle = \langle P_{ME}^{Tell}\rangle + \langle P_{ME}^{chir}\rangle. \quad (28)$$

At the same time, based on the above modeling of energy circulation *in the subwavelength domain* it can be argued that EM radiation cannot be a plane wave. It has a *vortex structure* of the fields with both spin and orbital angular momenta. For the energy balance in a such domain, $\langle W_{ME}^{Tell}\rangle$ can be considered as the potential and $\langle W_{ME}^{chir}\rangle$ as the kinetic parts of energy.

In their seminal paper [53], Tang and Cohen used Eq. (26) to show that a chiral molecule subjected to a monochromatic EM field generates an electric-dipole and a magnetic-dipole moments, and these moments are subjected to the mixed electric-magnetic dipole polarizability. They introduced a measure of chirality density, called optical chirality, as a *time-even pseudoscalar* that can be observed with circularly polarized EM plane waves. For time-harmonic fields, the optical chirality is expressed as $C = -\frac{\omega \varepsilon_0 \mu_0}{2} \mathrm{Im}(\vec{E} \cdot \vec{H}^*)$. They specifically emphasized that this is a measure of the *local* density of chirality. It was shown that enhanced chiral asymmetry can be achieved at the nodes of a specially constructed optical standing wave, which is composed by two counterpropagating circularly polarized plane waves with slightly different intensity. Small molecules localized to this region are predicted to show enhanced chiral asymmetry in their rate of excitation. Fig. 11 illustrates Tang and Cohen's model. It is supposed that for two counterpropagating circularly polarized plane waves with slightly different intensity we have a 2D region with enhanced chiral asymmetry. However, the interference of propagating waves is not actually a near-field structure. The standing wave constructed from these two plane waves corresponds to wavenumbers on the scale about 1/*r*. The chiral sample is viewed as a composition of electric and magnetic dipoles which are not coupled *locally* by quasistatic ME energy. Since in this case, we can analyze a thin sheet of chiral molecules, surface-plasmon structures can be effectively used for sensing [54]. In Ref. [55], the authors argue that the generation of "superchiral" nodes observed in experiments [54] when circularly polarized light is passed through a chiral film can be explained in terms of signal amplification arising from the well-known effect of surface plasmon amplification in systems fabricated with a metal substrate. Another important issue concerns the fact that the optical chirality density is a measure of only the spin of a photon. The orbital angular momentum of a photon is not taken into account [56].

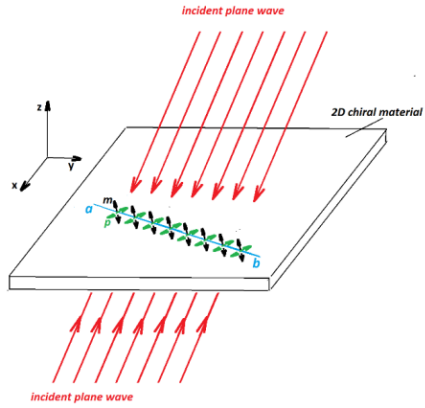

**Fig. 11.** Illustration of Tang and Cohen's model. For two counterpropagating circularly polarized plane waves with slightly different intensity we have a 2D region with enhanced chiral asymmetry. On the *xy* plane, the electric *p* and magnetic *m* dipole moments of chiral molecules oscillate without phase variation along the line *ab*. No real ME energy of interaction between *p* and *m* dipole moments is assumed. The imaginary part of the ME energy is observed due to the far-field EM radiation.

In our above consideration, we argued that when the effect of local coupling between electric and magnetic dipole excitations are introduced, both spatial and temporal inversion symmetry violations are necessary to satisfy the Poynting theorem. In the analysis [53], the time-reversal violation mechanism was not assumed. So, only an imaginal part of the ME energy is taken into account. It is supposed that optical chirality is a fundamental property of the electromagnetic field on an equal footing with energy density the Poynting theorem. Really, both quantities are quadratic in the fields. However, the energy density in Maxwell equations has a clear physical meaning as the energy density in the quasistatic limit [4, 33]. Although Tang and Cohen introduced optical chirality as a *local measure* of the chirality of the circularly polarized EM radiation, they did not consider this parameter to be meaningful in the *quasistatic limit*. In the continuity equation written for optical chirality $C$ and optical chirality flux $\vec{G}$

$$\frac{\partial C}{\partial t} + \frac{1}{\mu_0}\nabla \cdot \vec{G} = -\frac{1}{2}\left(\vec{j}\cdot\nabla\times\vec{E} + \vec{E}\cdot\nabla\times\vec{j}\right), \quad (29)$$

where $C \equiv \frac{\varepsilon_0}{2}\vec{E}\cdot\nabla\times\vec{E} + \frac{1}{2\mu_0}\vec{B}\cdot\nabla\times\vec{B}$ and $\vec{G} \equiv \frac{1}{2}\left[\vec{E}\times(\nabla\times\vec{B}) - \vec{B}\times(\nabla\times\vec{E})\right]$, we have the optical chirality expressed by *both* electric and magnetic field curls. Thus, this is a non-local (non-subwavelength) parameter which can only be detected due to the EM wave *propagation* process. In this connection, arises a question on the physical realizability of *point source* generation of superchiral fields [57, 58]. Generation of superchiral fields by such a "point source" should presume existence of orbital angular momentum of the EM fields. One cannot observe chiral near fields of a 3D-confined subwavelength sample based on the optical chirality parameter $C$.

Complementary, it is worth noting here that initially, in the study of the chiral light-matter interaction, the coupled dipole approximation was used for *far-field scattering* solutions [59 – 62]. For a system of two chiral molecules located in points $\vec{r}_a$, $\vec{r}_b$, the Hamiltonian of the molecule-field coupling is described as [62]

$$\mathcal{H}_{int} \propto \vec{p}(\vec{r}_a)\cdot\vec{E}^*(\vec{r}_a) + \vec{p}(\vec{r}_b)\cdot\vec{E}^*(\vec{r}_b) + \vec{m}(\vec{r}_a)\cdot\vec{B}^*(\vec{r}_a) + \vec{m}(\vec{r}_b)\cdot\vec{B}^*(\vec{r}_b). \quad (30)$$

With this form of $\mathcal{H}_{int}$, interactions at all distances are solely *mediated by the radiation field by exchange of photons*. Energy passes via the radiation field in left- and right-circularly polarized photons. For the free molecules, the free radiation field, and the molecule-field coupling, a total Hamiltonian $\mathcal{H} = \mathcal{H}_{mol} + \mathcal{H}_{rad} + \mathcal{H}_{int}$ does not contain a term of the ME interaction between the electric and magnetic dipoles.

## IV. ME META-ATOMS AND ME NEAR-FIELD PROBING OF CHIRALITY

The term "chiral" is traditionally used to denote structural chirality as a phenomenon without mirror symmetry. This defines the properties of electromagnetic waves that are commonly used for chirality discrimination. We argue that for local (deep subwavelength) determination of chirality properties, the engineered *near fields*, called the ME fields, should be characterized by broken *both* reflection and time-reversal symmetries. In such a case, the energy density in the vacuum near-field region of ME meta-atoms and probes for the chirality testing, is described as

$$\langle W \rangle = \frac{1}{2}\varepsilon_0\left|\vec{E}\right|^2 + \frac{1}{2}\mu_0\left|\vec{H}\right|^2 + \frac{1}{2}\sqrt{\varepsilon_0\mu_0}\,\mathrm{Re}\left(\vec{H}^*\cdot\vec{E}\right) + \frac{1}{2}\sqrt{\varepsilon_0\mu_0}\,\mathrm{Im}\left(\vec{H}^*\cdot\vec{E}\right). \quad (31)$$

Although this equation obviously follows from Eqs. (23), (24) and (27), it should be noted that the question of the near-field structure and quasistatic energy of coupled electric and magnetic dipoles in a 3D-confined subwavelength sample remains open. To answer this question, it is necessary to analyze the properties of magnetostatic (MS) and electrostatic (ES) oscillations in a 3D-confined structure.

To introduce the concept of optical chirality, Tang and Cohen [53] used Lipkin's "zilch" [63]. Let us consider this "zilch" as a combination of two terms:

$$C = F^{(E)} + F^{(H)}, \tag{32}$$

where

$$F^{(E)} \equiv \frac{\varepsilon_0}{2} \vec{E} \cdot \nabla \times \vec{E} \tag{33}$$

and

$$F^{(H)} \equiv \frac{1}{2\mu_0} \vec{B} \cdot \nabla \times \vec{B} \tag{34}$$

For time harmonic fields, we will define the real and imaginary parts of parameters $F^{(E)}$ and $F^{(H)}$ $\left( F^{(E,H)} = \left( F^{(E,H)} \right)' + i \left( F^{(E,H)} \right)'' \right)$ as follows:

$$\begin{aligned}
\left( F^{(E)} \right)' &= \frac{\varepsilon_0}{2} \operatorname{Im} \vec{E} \cdot \left( \nabla \times \vec{E} \right)^* = \frac{\omega \varepsilon_0 \mu_0}{2} \operatorname{Re} \left( \vec{E} \cdot \vec{H}^* \right) \\
\left( F^{(E)} \right)'' &= \frac{\varepsilon_0}{2} \operatorname{Re} \vec{E} \cdot \left( \nabla \times \vec{E} \right)^* = \frac{\omega \varepsilon_0 \mu_0}{2} \operatorname{Im} \left( \vec{E} \cdot \vec{H}^* \right)
\end{aligned} \tag{35}$$

and

$$\begin{aligned}
\left( F^{(H)} \right)' &= \frac{1}{2\mu_0} \operatorname{Im} \vec{B} \cdot \left( \nabla \times \vec{B} \right)^* = -\frac{\omega \varepsilon_0 \mu_0}{2} \operatorname{Re} \left( \vec{H} \cdot \vec{E}^* \right) \\
\left( F^{(H)} \right)'' &= \frac{1}{2\mu_0} \operatorname{Re} \vec{B} \cdot \left( \nabla \times \vec{B} \right)^* = -\frac{\omega \varepsilon_0 \mu_0}{2} \operatorname{Im} \left( \vec{H} \cdot \vec{E}^* \right).
\end{aligned} \tag{36}$$

It is obvious that for propagating EM waves, the introduction of two separate parameters $F^{(E)}$ and $F^{(H)}$ looks formal, having no physical meaning. However, it makes sense for quasistatic oscillations in the near-field zone. We have non-zero quantities $F^{(E)}$ and $F^{(H)}$ for magnetostatic ($\nabla \times \vec{H} = 0$) and electrostatic ($\nabla \times \vec{E} = 0$) oscillations, respectively. The MS and ES oscillations can be observed in a 3-D confined subwavelength sample of a magnetic insulator. An example of such a sample is the normally magnetized quasi-2D ferrite disk considered above. In the quasistatic resonances, a continuum approach is applied and quantum confinement effects are analyzed based on the concept that the MS function $\psi$ ($\vec{H} = -\vec{\nabla} \psi$) and the ES function $\phi$ ($\vec{E} = -\vec{\nabla} \phi$) are *scalar wave functions*. In spectral problem solutions, MS and ES wave functions play the role of order parameters for interacting dipoles. The coupling states of two, MS and ES, concurrent orders are considered as the ME states. While the quasistatic oscillations are not identified with the scales of individual atoms, they can be represented as dynamics of quasiparticles of the elementary dipole-dipole excitations. Scalar wave functions are the generating functions of the vector fields. It is worth noting that in a quasi-2D ferrite-disk resonator, dynamical electric polarization (ES resonances) is a by-product of the magnetization dynamics (MS resonances). The quantities $F^{(E)}$ and $F^{(H)}$ are called the helicity parameters. They constitute the real ("Tellegen") part of the ME energy. Together with this, the quasistatic oscillations are characterized by the power-flow vortices. They constitute the imaginary ("chiral") part of the ME energy [29, 43, 64]. This analysis can clarify the physical meaning of the energy density terms in Eq. (31).

Dynamical processes in a 3D-confined *subwavelength* open resonator do not experience a spatial gradient in the electric and magnetic fields, only a temporal change. For such point scatterers as electric or magnetic dipoles, the near-field structure and quasistatic energy of are well known. The potential energies of the electric and magnetic dipoles in external electric and magnetic fields are, respectively, $W_E = -\vec{p}\cdot\vec{E}$ and $U_M = -\vec{m}\cdot\vec{B}$. In the near-field zone, the fields of the electric and magnetic dipoles are, respectively, dominantly electric and dominantly magnetic fields. Since in the quasistatic limit, the magnetic field of the electric dipole and the electric field of the magnetic dipole, respectively, disappear, we can say that the near-field zones of subwavelength electric and subwavelength magnetic dipoles *extend to infinity* [33]. Obviously, from the point of view of classical electrodynamics, coupling electric and magnetic dipoles in the near-field zone is impossible. To analyze a ME meta-atom as a resonant element which size is much smaller than the EM wavelength, we have to use the quantum description. This description is not based on the *microscopic*-scale quantization (based on the electron scalar-wave function), but on the *mesoscopic*-scale quantization (based on the quasistatic scalar-wave function). As we noted above, quasistatic oscillations in a 3D-confined subwavelength sample are considers in nonequalities $\lambda_{em} \gg \lambda_{qs} \gg \lambda_{at}$. These oscillations are viewed as processes occurring in the quasistatic limit $\lambda_{em} \to \infty$. The energy transitions between atom states are known as Rabi oscillations. Quasistatic modes in a quasi-2D ferrite disk have the energy eigenstate spectrum. At the external microwave radiation, these spectral properties occur due to topological effects. One observes quantized states of potential energy and quantized power-flow vortices. This effect of mesoscopic-scale quantization is described based on the quasi-magnetostatic and quasi-electrostatic scalar wave functions [43, 64 – 67]. Such quantized states are well observed in the cavity ME effect. This effect demonstrates the relationship between the *ME meta-atom* and the *ME photon*. Since the near-field zone corresponds to the quasistatic limit, spectra of quasistatic oscillations extend to infinity. Due to the ME energy, the entire EM continuum becomes strongly modified in the far-field regions. This modified EM radiation, called ME photon, is rotationally supperradiant radiation.

As we discussed above, using the circular dichroism effects for chirality discrimination gives us the far-field sensing mechanism in optics. While considering chirality as the phenomenon fundamentally related to the ME response, we argue that in engineering enantioselective near fields for local (subwavelength) chirality discrimination, both spatial and temporal inversion symmetry violations are necessary. The fact that in a near-field region of EM radiation, one can observe both violation of space and time-reversal symmetry of chiral molecules was affirmed by experiments in Ref. [68]. In this work, efficient enantio-discrimination arises from the possibility of applying ME probing fields to near-field microwave spectroscopy. Together with a theoretical model suggested in [68], the most appropriate theoretical explanation of this effect can be made based on the chiral induced spin selectivity (CISS) scheme describing the charge and spin polarization in chiral molecules. When electrons move through a chiral molecule, the resulting current may become spin polarized [69, 70]. A sketch illustrating the CISS model for the analysis of microwave experiments [68] is shown in Fig. 12. A capsule of aqueous D- and L-glucose solutions is placed on a quasi-2D ferrite disk with quasistatic resonances. Magnetization dynamics inside a ferrite disk is characterized by both spin and orbital angular momentums. In a glucose sample, the induced electric-dipole polarization accomplish rotation with both spin and orbital angular momentums. A bias magnetic field is directed normally to the disk plane. For opposite directions of a bias magnetic field, one has opposite rotations (of both orbital and spin momenta) of the fields inside a ferrite and a sample. A chiral molecule is presented schematically as a coil. The molecule has a preferred orientation of the electron spin after charge polarization. This orientation of the electron spin depends on the handedness of the molecule. As electrons move through a chiral molecule, an effective magnetic field is generated. A preference exists for electrons with one magnetic moment direction to pass through the chiral molecule. In experiment [68], the orbital rotation of the fields causes the coupling of a very large number of chiral molecules, with a corresponding increase in the strength of the oscillator. We observe exceptionally large quantum-coherent coupling: chiral molecules – magnons – microwave photons. Quasistatic resonances in a ferrite disk "feel" and "discriminate" molecular chirality. In the experiment, the enantiomer-dependent parameters of the scattering matrix are analyzed for two opposite directions of a bias magnetic field for the D- and L-glucose solutions.

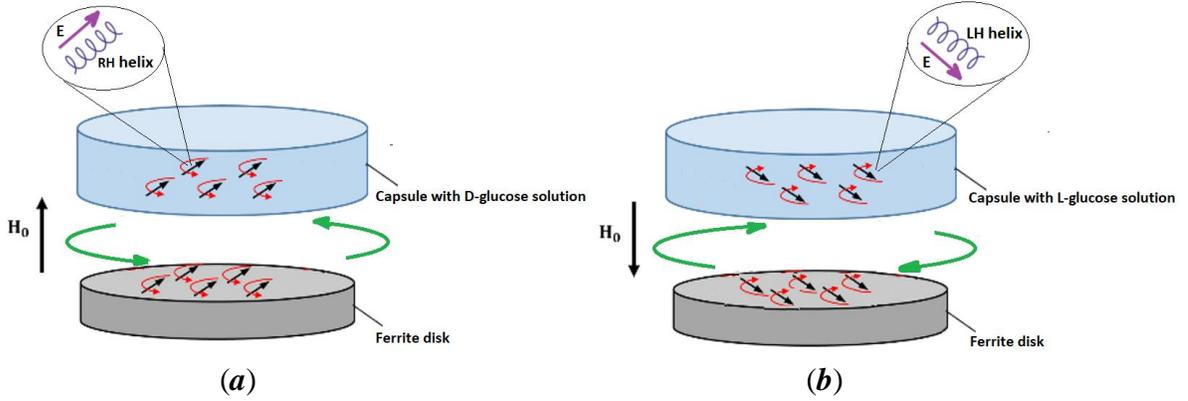

(*a*)           (*b*)

**Fig. 12.** A model illustration of the experiment in Ref. [68]. Effective enantio-discrimination results from the application of ME probing fields to microwave near-field spectroscopy. A quasi-2D ferrite disk with quasistatic resonances is loaded by a cylindrical capsule with aqueous D- or L-glucose solutions. The dynamics of the magnetization inside the ferrite disk and the dynamics of the electric dipole polarization inside the sample are shown schematically by rotating black arrows. Chiral molecules are presented schematically as right-hand and left-hand helices. The electric field **E** induces a spin-selective electron displacement. Quasistatic resonances in a ferrite disk "feel" and "discriminate" molecular chirality. ME responses, characterized by the violation of both parity (*P*) and time-reversal (*T*) symmetry, are analyzed by examining the scattering matrix parameters. The scattering matrix parameter $S_{ij}$ detected in the cases (*a*) and (*b*) is the same: $\left(S_{ij}^{H_0\uparrow}\right)^{(D)} = \left(S_{ij}^{H_0\downarrow}\right)^{(L)}$. This means that the broken *P* and *T* symmetries are preserved in the combined *PT* symmetry.

## V. CONCLUSION

Electromagnetism is characterized by the effect of coupling between the polar (electric field) and axial (magnetic field) vectors on the scale of the EM wavelength. In the light-matter interaction problems, magnetoelectricity, on the contrary, is considered as the effect of the coupling between the polar (electric dipole) and axial (magnetic dipole) vectors in the near-field zone. At the same time, it is known that since in the deep subwavelength region, the fields of the electric and magnetic dipoles are, respectively, dominantly electric or dominantly magnetic, ME coupling of these dipoles is impossible. Thus, the potential energies of the electric and magnetic dipoles are not mutually converted. In other words, from the point of view of classical electrodynamics, there is no ME energy in such a region. All this makes ME electrodynamics a very nontrivial task.

In areas of research called electromagnetic chirality [71 – 75] and, in a general case, electromagnetic bianisotropy [48], it is assumed that under EM radiation, material constituents (such as molecules or meta-atoms) acquire a ME response. In the dipole approximation, this response is considered as being associated with local cross-coupling between the electric and magnetic polarizations. This model essentially consists of neglecting the spatial variation of the electromagnetic field over the molecule (meta-atom) – instead, one uses the field at the location of the element. In metamaterial electrodynamics, meta-atoms are considered as subwavelength resonant elements. It is assumed that these elements can be viewed as small dipole antennas. The region closest to such a small antenna forms a reactive near-field region. Thus, in this case, no response with the real ME energy of the local coupling of the electric and magnetic fields can be assumed.

In our paper, we argue that real ME energy can occur when ME fields in a singular subwavelength domain are characterized by a violation of both the symmetry of time reversal and spatial reflection. These ME fields are characterized by the eigenstate spectrum of ME energy. The energy eigenstate problem can be correctly formulated as the quasistatic resonance problem in a subwavelength domain [29, 43, 64, 65]. This distinguishes our concept from the concept of ME energy density proposed in Ref. [76]. In the latter case, we have actually nonlocal effects of the interaction of two counterpropagating circularly polarized plane waves. We also argue that ME near fields are true enantioselective fields for describing local (subwavelength) properties of chiral light-matter interactions. While enantioselective fields defined from time-even pseudoscalar optical chirality (as a result of interaction of two counterpropagating plane waves) [53], are not actually near-field structures.

Due to the ME energy, the entire EM continuum is strongly modified both in the near-field and far-field regions. The fundamental study of the interaction of a resonant point ME scatterer with EM radiation constitutes an important area of new research related to the concept of quantum spacetime [77].